\documentclass{ws-procs9x6-cpt19}
\begin{document}

\newcommand{\refeq}[1]{(\ref{#1})}
\def\etal {{\it et al.}}
\newcommand{\diff}[1]{\text{d}#1}
\newcommand{\Lie}{\mathcal{L}}
\newcommand{\D}{\text{D}}
\newcommand*{\diag}{\operatorname{diag}}

\title{Symmetries in the SME gravity sector: A study in the first-order formalism}

\author{Y.\ Bonder$^1$ and C.\ Corral$^{1,2}$}

\address{$^1$Instituto de Ciencias Nucleares, Universidad Nacional Aut\'onoma de M\'exico,\\
Circuito Exterior s/n Ciudad Universitaria, Ciudad de M\'exico, 04510, M\'exico}

\address{$^2$Universidad Andr\'es Bello, Departamento de Ciencias F\'isicas, \\ Facultad de Ciencias Exactas, Sazi\'e 2212, Santiago, 8370136, Chile}

\begin{abstract}
A method to find the symmetries of a theory in the first order formalism of gravity is presented. This method is applied to the minimal gravity sector of the Standard Model Extension. It is argued that no inconsistencies arise when Lorentz violation is explicit and the relation between Lorentz violation and invariance under (active) diffeomorphisms is clearly exposed.
\end{abstract}

\bodymatter

The Standard Model Extension\cite{SME} (SME) is a framework to parameterize all possible violations of local Lorentz invariance. It has a gravitational sector\cite{SMEgrav} where the fields describing the spacetime geometry are coupled to the SME coefficients. Here, only explicit Lorentz violation is considered where the SME coefficients are nondynamical.

On the other hand, the first-order formalism of gravity\cite{FOG} has the vierbein and an independent Lorentz connection as the dynamical geometrical variables. The mathematical framework is that of differential $p$-forms, i.e., totally antisymmetric tensors of rank $(0,p)$, $p\leq 4$. In fact, the vierbein $e^a$ is an $\mathfrak{so}(1,3)$ valued $1$-form (Latin indexes are Lorentz indexes; spacetime indexes are omitted) and the Lorentz connection is a $1$-form $\omega^{ab} = -\omega^{ba}$.

The basic operations of this framework are the wedge product $\wedge$, the inner product with respect to the vector field $\xi$, $i_\xi$, and the exterior derivative $\diff{}$. Basically, $\wedge$ is a tensor product whose result is antisymmetrized, $i_\xi$ saturates the $p$-form with $\xi$, thus reducing the rank of the form by one, and $\diff{}$ is an antisymmetrized derivative using any torsionless derivative operator that raises the rank of the form by one. In addition, the Minkowski metric $\eta_{ab}= \diag(-1,1,1,1)$ and its inverse $\eta^{ab}$ are used to lower and raise Lorentz indexes and the summation convention over repeated indexes is assumed. The conventions that are used are those of Ref.\ \refcite{Nakahara}.

The central equations of this formalism are Cartan's structure equations for the curvature and torsion $2$-forms, $R^{ab}$ and $T^{a}$:
\begin{equation}
R^{ab}= \diff{ \omega^{ab}} + {\omega^a}_c \wedge \omega^{cb},\quad 
T^a= \diff{ e^a}+ {\omega^a}_b \wedge e^b.\label{Cartan2}
\end{equation}
Moreover, the Lorentz connection can be used to define a covariant exterior derivative $\D$. This is done by taking the conventional exterior derivative and adding (subtracting) a term for each superscripted (subscripted) Lorentz index. By definition, $\D$ is Lorentz covariant. Using that $\diff{}^2=0$, it is easy to show that $ \D R^{ab}=0$ and $ \D T^{a}=R^{a}{}_{b}\wedge e^b$, which are the Bianchi identities.

The main advantages of the first-order formalism are that it is suitable for integration and efficient to perform action variations. In addition, considers a more general connection that is \textit{a priori} torsion full; to recover the results of the metric formalism one need to set torsion to zero consistently. Moreover, the Lie derivative along a vector field $\xi$, when acting on a $p$-form $\theta$, simply becomes $\Lie_\xi \theta= i_\xi \diff{ \theta} + \diff{ i_\xi \theta}$; this is known as Cartan's formula.\cite{Nakahara}

At this point attention is set on applications of the first-order formalism in the SME. The minimal part of the gravitational sector of the SME (mgSME) has Lorentz violating terms modifying the action of General Relativity (GR) with no additional derivatives. In vacuum and in the first-order formalism the corresponding action is\cite{Sym}
\begin{equation}
 S_{\rm mgSME}[\omega^{ab},e^a] = \frac{1}{2\kappa}\int\left(\epsilon_{abcd} + k_{abcd} \right)R^{ab}\wedge e^c\wedge e^d,
 \end{equation}
where $\kappa$ is the gravitational coupling constant, $\epsilon_{abcd}$ is the totally antisymmetric Lorentz tensor and $k_{abcd}$ is a nondynamical $0$-form that plays the role of the SME coefficients. The term proportional to $\epsilon_{abcd}$ corresponds to the Einstein--Hilbert action in the presence of torsion (or the Einstein--Cartan action) without a cosmological constant. Interestingly, due to the presence of torsion, there are more coefficients than in the conventional mgSME. In particular, $k_{abcd}$ is such that $k_{abcd} = k_{[ab][cd]} $ but $k_{abcd}\neq k_{cdab}$, reflecting the fact that the Ricci tensor is not symmetric (squared brackets denote antisymmetrization with a factor $1/2$).

An arbitrary variation of compact support yields
\begin{eqnarray}
 \delta S_{\rm mgSME} &=&\int\left(  \delta\omega^{ab}\wedge\mathcal{E}_{ab}+\delta e^a\wedge\mathcal{E}_a \right) ,\label{actionvar}\\
 \mathcal{E}_{ab} &=& \frac{1}{\kappa}\left(\epsilon_{abcd} + k_{abcd} \right)T^c\wedge e^d + \frac{1}{2\kappa}\D {k_{abcd}}\wedge e^c\wedge e^d,\label{eom1}\\
  \mathcal{E}_a &=&\frac{1}{\kappa}\left(\epsilon_{abcd} + k_{abcd} \right)R^{bc}\wedge e^d.\label{eom2}
\end{eqnarray}
Clearly, $\mathcal{E}_{ab}=0=\mathcal{E}_{a}$ are the equations of motion (EOM), thus, $\mathcal{E}_{ab}$ and $\mathcal{E}_{a}$ are called EOM throughout the text. However, the symmetries must be studied off shell and the EOM are not assumed to vanish. The covariant exterior derivatives of the EOM can be casted into the form
\begin{eqnarray}
\D{\mathcal{E}_{ab}} &=& e_{[a}\wedge\mathcal{E}_{b]} + \frac{1}{\kappa}\bigg(-2 k_{[a|cde|}  R^{cd}\wedge e_{b]}\wedge e^e\nonumber\\
 & &\quad+ k_{abcd}R^{c}{}_e\wedge e^e\wedge e^d -k_{[a|ecd|} R^{e}{}_{b]}\wedge e^c\wedge e^d \bigg),\label{DE1}\\
\D{\mathcal{E}_a} &=& i_a T^b\wedge\mathcal{E}_b + i_a R^{bc}\wedge\mathcal{E}_{bc} - \frac{1}{\kappa} i_a\D{k_{lbcd}}\wedge R^{bc}\wedge e^l\wedge e^d, \label{DE2} 
\end{eqnarray}
where $i_a$ is such that $i_\xi = \xi^a i_a$. Importantly, all term that cannot be written as some tensor contracted with the EOM are called symmetry breaking terms (SBT), and, in this case, the SBT are those terms with a $k_{abcd}$.

The symmetries and the arising conditions can be read off from Eqs.\ \eqref{DE1} and \eqref{DE2} using the following method (see Ref.\ \refcite{CQG}): Step 1, multiply these equations by the `gauge parameters,' namely, by Lorentz-valued $0$-forms of compact support $\lambda^{ab}=-\lambda^{ba}$ and $\xi^a$, respectively. Step 2, integrate over spacetime and use the Leibniz rule to convert $\lambda^{ab}\D \mathcal{E}_{ab} $ and $\xi^{a}\D \mathcal{E}_{a} $ to $\D \lambda^{ab}\wedge \mathcal{E}_{ab} $ and $\D \xi^{a} \wedge\mathcal{E}_{a} $. The resulting equations are
\begin{eqnarray}
0&=&\int \bigg[\D\lambda^{ab}\wedge \mathcal{E}_{ab} -\lambda^{ab}e_{b}\wedge\mathcal{E}_{a} + \frac{\lambda^{ab}}{\kappa}\bigg(-2 k_{acde}  R^{cd}\wedge e_{b}\wedge e^e\nonumber\\
 & &\quad+ k_{abcd}R^{c}{}_e\wedge e^e\wedge e^d -k_{aecd} R^{e}{}_{b}\wedge e^c\wedge e^d \bigg)\bigg],\label{intDE1}\\
0&=&\int \bigg[ i_\xi R^{ab}\wedge\mathcal{E}_{ab}+\left(D \xi^a+i_\xi T^a\right)\wedge\mathcal{E}_a  - \frac{1}{\kappa} i_\xi \D{k_{abcd}}\wedge R^{bc}\wedge e^a\wedge e^d\bigg], \label{intDE2} 
\end{eqnarray}
Step 3, verify if there exists a nontrivial gauge parameter $\tilde{\lambda}^{ab}$ ($\tilde{\xi}^a$) such that the SBT in Eq.\ \eqref{intDE1} (Eq.\ \eqref{intDE2}) vanish. If this occurs, this equation takes the form of the action variation \eqref{actionvar} and, by comparison, it is possible to read off the field transformations: $\delta \omega^{ab}=\D \tilde{\lambda}^{ab}$ and $\delta e^{a}=-\tilde{\lambda}^{ab}e_{b}$ ($\delta \omega^{ab}=i_{\tilde{\xi}} R^{ab}$ and $\delta e^{a}=\D \tilde{\xi}^{a}+ i_{\tilde{\xi}} T^{a}$). Conversely, if there are no gauge parameters such that the SBT vanish, then the theory has no symmetries.

Note that the transformation laws for the vierbein and the Lorentz connection obtained from Eq.\ \eqref{intDE1} coincide with their well-known Lorentz transformations. Since, for an arbitrary $k_{abcd}$, there is no nontrivial $\tilde{\lambda}^{ab}$ such that the SBT vanish, it is possible to conclude that this symmetry is completely broken in the mgSME. On the other hand, the transformations arising from Eq.\ \eqref{intDE2} are not diffeomorphisms (Diff), as one could naively expect, but a covariant Diff in which, in Cartan's formula, $\diff{}$ is replaced by $\D $. This suggests that the fundamental symmetries of the theories under consideration, including GR, are not the conventional Diff but their covariant version. Again, for an arbitrary $k_{abcd}$ there is no $\tilde{\xi}^{a}$ for which the SBT vanish, therefore, the mgSME breaks this symmetry. Notice, however, that one can have invariance under the covariant Diff (e.g., $\D k_{abcd}=0$), but the theory can still break the conventional Diff invariance since the Diff are a combination of a covariant Diff and a Lorentz transformation. Another interesting example is the unimodular version of Einstein--Cartan theory\cite{CQG,Unimod}, which is Lorentz invariant but is not invariant under all Diff.

In conclusion, a rigorous method to find the symmetries in a particular theory is presented. With this method, it can be shown\cite{Sym} that no inconsistencies arise between the Bianchi identities and a nondynamical $k_{abcd}$. At most, the conservation laws impose restrictions on $k_{abcd}$, which goes against the typical SME work hypothesis. Another lesson from this analysis is that the interplay of Lorentz and Diff violation is richer than is usually considered. Other possible applications of the first-order formalism in the context of explicit Lorentz violation can include the construction of the Hamiltonian by following the method of Ref.\ \refcite{Hamilt} and the construction of nonminimal terms using the fact that $4$-forms are the natural objects in the action, where the Hodge dual will certainly play a role.

\section*{Acknowledgments}
This work was supported by UNAM-DGAPA-PAPIIT Grant IA101818.


\begin{thebibliography}{xx}

\bibitem{SME}
D.\ Colladay and V.A.\ Kosteleck\'y, Phys.\ Rev.\ D \textbf{55}, 6760 (1997); Phys.\ Rev.\ D \textbf{58}, 116002 (1998).

\bibitem{SMEgrav}
V.A.\ Kosteleck\'y, Phys.\ Rev.\ D \textbf{69}, 105009 (2004); Y.\ Bonder,  Phys.\ Rev.\ D \textbf{91}, 125002 (2015); Y.\ Bonder and G.\ Le\'on, Phys.\ Rev.\ D \textbf{96}, 044036 (2017).

\bibitem{FOG}
M.\ Blagojevi\'c and F.W.\ Hehl, eds., \textit{Gauge Theories of Gravitation}, World Scientific Publishing, 2013.

\bibitem{Nakahara}
M.\ Nakahara, \textit{Geometry, Topology and Physics}, Taylor \& Francis, 2016.

\bibitem{Sym}
Y.\ Bonder and C.\ Corral, Symmetry \textbf{10}, 433 (2018).

\bibitem{CQG}
Y.\ Bonder and C.\ Corral, Phys.\ Rev.\ D \textbf{97}, 084001 (2018).

\bibitem{Unimod}
C.\ Corral and Y.\ Bonder, Class. Quantum Grav. \textbf{36}, 045002 (2019).

\bibitem{Hamilt}
J.M.\ Nester, Mod. Phys. Let. A \textbf{6}, 2655 (1991); C.M.\ Chen, J.M.\ Nester, and R.S.\ Tung, Int. J. Mod. Phys. D \textbf{24}, 1530026 (2015).
\end{thebibliography}
\end{document}